\documentclass[twocolumn]{aastex701}

\usepackage{threeparttable}
\usepackage{verbatim}  
\usepackage{graphicx}                    % For eps figures, newer & more powerfull
\usepackage{color}                       % For color text: \color command
\usepackage{breakurl}                         % For breaking URLs easily trough lines
\usepackage{rotating}
\usepackage{hyperref}
\usepackage{siunitx}
\usepackage{graphicx}
\usepackage{amsmath}
\usepackage{threeparttable}
\usepackage{subcaption}
\usepackage{ltablex} 
\usepackage{threeparttablex} % A special version of threeparttable built for long tables

\newcommand{\spwea}{Space Weather}

\begin{document}

\title{Deriving Interplanetary Sector Boundary Crossings using Potential Field Source Surface Modeling}

\author[orcid=0000-0002-2655-2108,sname='SGH']{Stephan~G.~Heinemann}
\affiliation{Department of Physics, University of Helsinki, Finland}
\affiliation{Institute of Physics, University of Graz, Austria}
\email[show]{stephan.heinemann@hmail.at}  

\author[orcid=0000-0003-4867-7558, sname='MT']{Manuela~Temmer} 
\affiliation{Institute of Physics, University of Graz, Austria}
\email{manuela.temmer@uni-graz.at}

\author[0000-0003-1175-7124, sname='JP']{Jens~Pomoell} 
\affiliation{Department of Physics, University of Helsinki, Finland}
\email{jens.pomoell@helsinki.fi}

\author[orcid=0000-0003-2061-2453, sname='MO']{Mathew~Owens} 
\affiliation{Department of Meteorology, University of Reading, Reading, UK,}
\email{m.j.owens@reading.ac.uk}

\author[orcid=0000-0002-4489-8073, sname='EK']{Emilia~Kilpua} 
\affiliation{Department of Physics, University of Helsinki, Finland}
\email{emilia.kilpua@helsinki.fi}

\author[orcid=0000-0001-9574-339X, sname='ML']{Matti~ala-Lahti} 
\affiliation{Department of Physics, University of Helsinki, Finland}
\email{matti.ala-lahti@helsinki.fi}

\author[orcid=0000-0002-5349-6308, sname='AW']{Andreas~Wagner} 
\affiliation{Department of Physics, University of Helsinki, Finland}
\email{andreas.wagner@helsinki.fi}

\begin{abstract}
The Heliospheric Current Sheet (HCS) and associated interplanetary sector boundaries fundamentally structure the solar wind environment, modulating coronal mass ejection (CME) evolution and solar energetic particle transport. In this study, we evaluate the capability of a static Potential Field Source Surface (PFSS; $R_{\mathrm{ss}}=2.5\,R_{\odot}$) coronal model to directly map interplanetary sector boundary crossings from the Sun to 1~au without dynamic heliospheric propagation models. We compare ballistically transported HCS crossings derived from GONG-ADAPT and SDO/HMI magnetic maps against sector boundary crossing observations from the SBCross catalog. PFSS modeling reproduces the large-scale interplanetary magnetic field topology well ($J\approx0.50$), achieving hit rates (Probability of Detection; POD) of 0.64 for GONG-ADAPT and 0.76 for HMI. Carrington synoptic maps outperform flux-transport modeled maps, suggesting that the short-term observational noise from data assimilation degrades topological matching more than map ``aging''. Predictive skill shows a strong solar cycle dependence, performing significantly better during solar maxima ($\mathrm{POD}=0.70-0.81$) than solar minima ($\mathrm{POD}=0.56-0.67$), attributed to more distinct crossings due to higher HCS inclination angles during maximum. Matched events exhibit travel times close to ballistic estimates of 4.8-4.9~days with an average timing difference centered near zero ($\pm1.1$~days) and 90\% of events within $\pm1.8$~days. The hit rate of HCS structures that interact with CMEs does not degrade  ($\mathrm{POD}=0.67-0.83$), despite propagating $\approx7$~hours faster on average, implying that the CME interaction causes only minor structural changes to the HCS. Larger timing differences ($>24$~hours) are associated with flatter current sheets. Overall, static PFSS extrapolations offer a computationally inexpensive, effective approach for mapping global sector boundary structure.
\end{abstract}



\section{Introduction} \label{s:intro}

% intro

% intro to sector boundaries

Outward expansion of magnetized coronal plasma, known as the solar wind, continuously resupplies the surrounding heliosphere \citep{Parker1961}. Embedded in this outflow is the Heliospheric Current Sheet \citep[HCS; e.g.,][]{Svalgaard1976, Wilcox1980, 2001Smith_HCS}, which acts as surface that separates regions of opposing magnetic polarity across interplanetary space. At 1\,au, this thin current layer (about $10,000~\text{km}$ thick) features a low magnetic field magnitude and supports an electrical current density on the order of $10^{-10}~\text{A m}^{-2}$ \citep{Winterhalter1994, 2001Smith_HCS, Eastwood2002}. The inclination between the Sun's magnetic dipole axis and the rotational axis, coupled with high order magnetic moments, impart a complex, three-dimensional wavy fold onto the HCS, a spatial geometry frequently described as the ``ballerina skirt'' \citep{Alfven1977}. This structural complexity is closely tied to the solar cycle: the evolving spatial distribution of coronal holes and open versus closed magnetic field regions gives rise to a dynamic, warped geometry as the Sun rotates. Consequently, any point in interplanetary space, such as a spacecraft or planet, periodically crosses the HCS, observing rapid magnetic field reversals and associated sector boundary crossings \citep{2001Smith_HCS}. \\ %This interface is enveloped by the Heliospheric Plasma Sheet \citep[HPS;][]{Burlaga1990}. The HPS displays signatures including elevated plasma $\beta$ and proton concentrations, paired with reduced magnetic field strengths, lower thermal temperatures, and a depleted relative helium content ($\mathrm{He/H}$) \citep{Gosling1981, Borrini1981, Winterhalter1994}. \\

To reconstruct interplanetary magnetic field polarity and associated polarity reversals, numerical models incorporate photospheric magnetic observations as boundary conditions. Typically, global photospheric magnetic maps are given as input to a coronal model such as a Potential Field Source Surface  \citep[PFSS;][]{Altschuler1969} or coronal MagnetoHydroDynamic (MHD) model \citep[e.g., MAS, COCONUT][]{mikic99a,Perri2022_COCONUT} to infer the HCS topology at a height of $R \approx 2.5\,R_\odot$. The coronal model output (specifically $B_r$) is then used as input to a heliospheric model e.g., empirical \citep[WSA;][]{wang90,Arge2000}, 1D hydrodynamic \citep[HUXt][]{Barnard2022} or 3D MHD codes such as ENLIL, MAS, or EUHFORIA to simulate the outward solar wind propagation \citep{1999enlil, 2006Riley, pomoell2018}. \\

\cite{Jian2015_CCMC} reported that solar wind models hosted at NASA's Community Coordinated Modeling Center (CCMC) correctly predict magnetic sector polarities with $70\%\text{--}90\%$ accuracy in times series at $1\text{~au}$, with timing errors of one to four days \citep[see also][]{MacNeice2009}. Subsequent empirical and numerical investigations have systematically evaluated the precision of different coronal-heliospheric model combinations \citep{Riley2012, Owens2013}. Benchmark studies comparing kinematic models (such as WSA combined with ENLIL or HUXt) against MHD formulations demonstrate that while 3D MHD captures fine-scale non-radial flow deflections near stream interaction regions, reduced-physics and kinematic models offer comparable sector-reversal skill at a fraction of the computational cost \citep{Owens2020, Barnard2022}. Overall, operational forecasts reliably reconstruct the global sector structure with mean temporal uncertainties in arrival time of $\Delta t \approx \pm 0.5\text{--}1.5\text{~days}$ relative to observations under quiet solar conditions \citep{Riley2012, Jian2015_CCMC}. \\

In addition to global topological mapping, local measurements of the HCS inclination provide boundary constraints for heliospheric modeling. Under quiet solar wind conditions, localized in-situ observations of the HCS orientation at $1\,\text{au}$ align closely with PFSS model predictions \citep{Klein1980, Burton1994, Peng2017}. However, stream interaction regions compress and distort the local current sheet, yielding an inclination steeper than the relatively flat HCS predicted by stationary coronal models; this dynamic steepening becomes pronounced at heliocentric distances $R > 2\,\text{au}$ \citep[e.g,][]{Pizzo1991,Pizzo1994}.  Coronal Mass Ejections (CMEs) further perturb this interface, as they frequently embed directly within the HCS structure rather than deflecting around it \citep{Crooker1993,Crooker1996,Forsyth1997,Temmer2026}. Although the global HCS re-establishes its configuration post-passage \citep{Zhao1996, Blanco2011}, the exact timescales governing this post-transient relaxation remain unquantified.\\

The spatial alignment of the HCS and the surrounding plasma environment fundamentally governs CME dynamic evolution and geoeffectiveness. Fast CMEs ($v > 800\,\text{km\,s}^{-1}$) overtaking the slow HCS structure ($v \approx 300\text{--}400\,\text{km\,s}^{-1}$) undergo dynamic compression, flattening at the CME front, locally amplifying the plasma density, plasma $\beta = \frac{p_{\text{thermal}}}{p_{\text{mag}}}$ (the ratio of thermal pressure to magnetic pressure), and southward magnetic field ($B_z$) at the interface \citep{Liewer2011,Liu2012,Temmer2025}. Upon impacting the magnetosphere, these compressed structures may deliver dynamic pressure spikes and sustained negative $B_z$, that regularly drive intense multi-phase geomagnetic storms with $Dst$ drops exceeding those from isolated CMEs \citep{Kataoka2015, Temmer2026}. In the corona, magnetic pressure gradients can deflect CMEs by $10^\circ\text{--}30^\circ$ toward the HCS, introducing arrival time errors up to $\Delta t \approx \pm 24\,\text{hr}$ and causing unexpected geomagnetic storms when high-latitude events are redirected toward the ecliptic \citep{Isavnin2014,Kay2015a,Winslowetal2016,wang2020deflection}. Current sheets also modulate Solar Energetic Particle (SEP) transport, acting as barriers or drift channels that significantly alter particle flux profiles \citep{Battarbee2017, Liou2024}.\\

In this study, we evaluate the capability of pure PFSS modeling for identifying HCS crossings at the source surface and evaluate the match of in-situ sector boundary crossings at 1 au. Instead of running  dynamic propagation models, we focus on isolating how well static magnetic extrapolations perform on their own. We take modeled polarity HCS crossings, transport them to 1 au, and evaluated them against a 20-year catalog of in-situ data to derive hit-miss statistic, timing precision, and solar cycle dependence. Finally, we investigate whether transient CMEs systematically alter how well these static models match in-situ events.

\section{Methodology}\label{s:methods}

\begin{figure}[htbp]
   \centering
   % Panel 1 (Top)
   \begin{subfigure}[b]{1\linewidth}
       \centering
       \includegraphics[width=\linewidth]{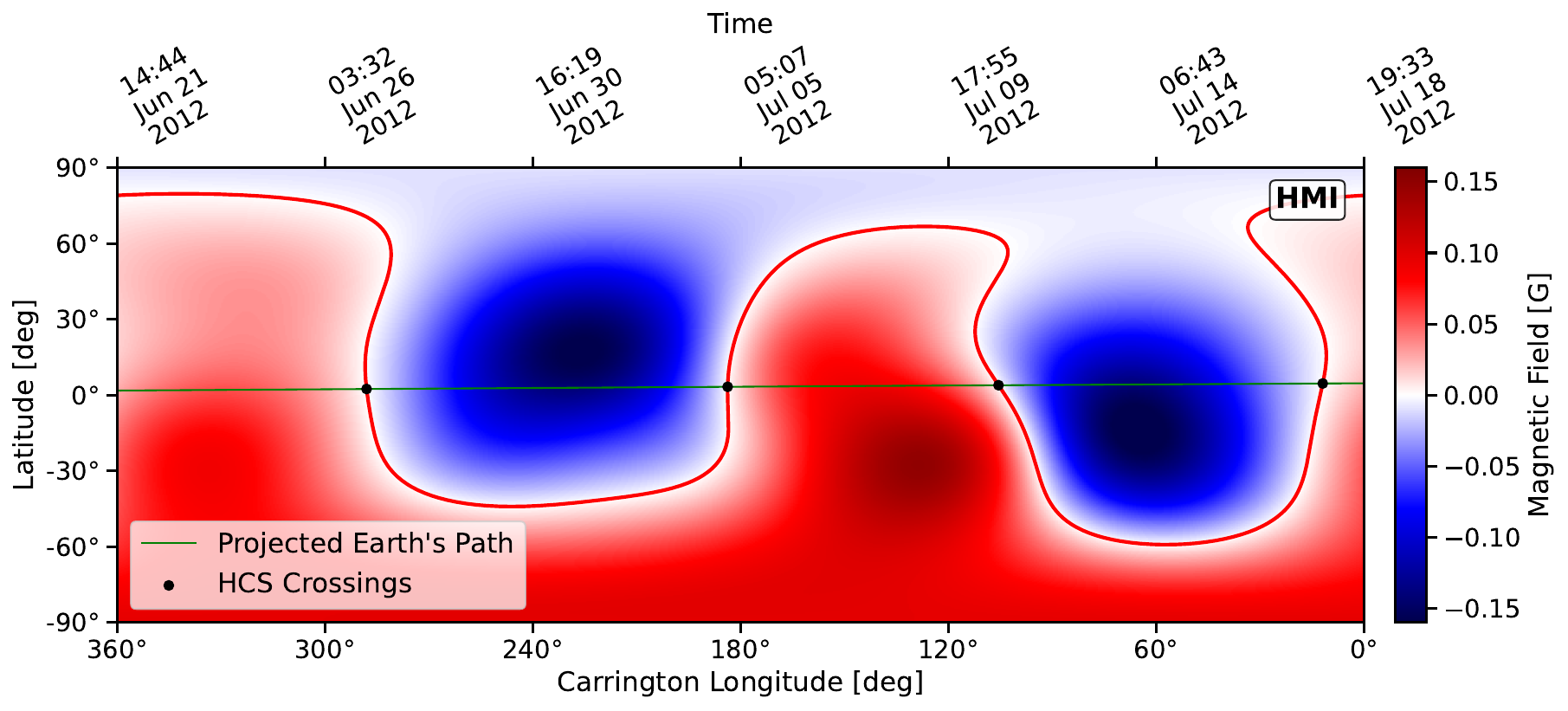}
       \phantomcaption % Generates subfigure counter for \ref{fig:pfss_a} without printing a text caption
       \label{fig:pfss_a}
   \end{subfigure}
   
   \vspace{-0.34cm} % Adjust space between panels if needed
   
   % Panel 2 (Bottom)
   \begin{subfigure}[b]{1\linewidth}
       \centering
       \includegraphics[width=\linewidth]{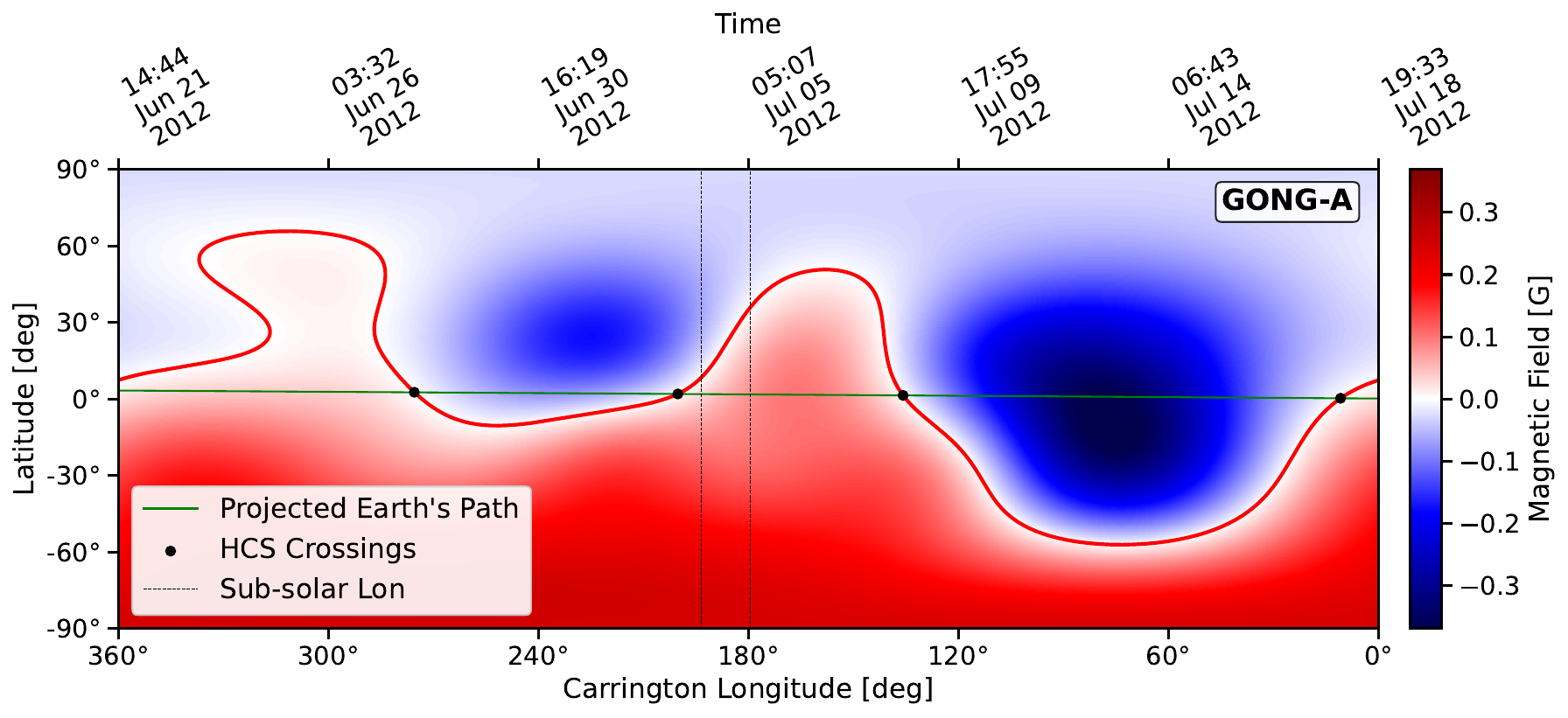}
       \phantomcaption % Generates subfigure counter for \ref{fig:pfss_b} without printing a text caption
       \label{fig:pfss_b}
   \end{subfigure}
   
    \caption{PFSS synoptic maps at the source surface ($R_{ss} = 2.5\,R_\odot$) derived from HMI data for Carrington Rotation 2125 (top) and daily GONG-A data for 2012-07-05 06:00 (bottom). Both panels display the radial magnetic field strength, the HCS (red line), the projected trajectory of Earth (green line), and modeled HCS crossings (black dots). The bottom panel additionally indicates the sub-solar longitude (vertical dotted lines) used to determine HCS crossings in GONG-A data.
    }
   \label{fig:pfss}
\end{figure}

\begin{figure*}[h]
   \centering
   \includegraphics[width=1\linewidth]{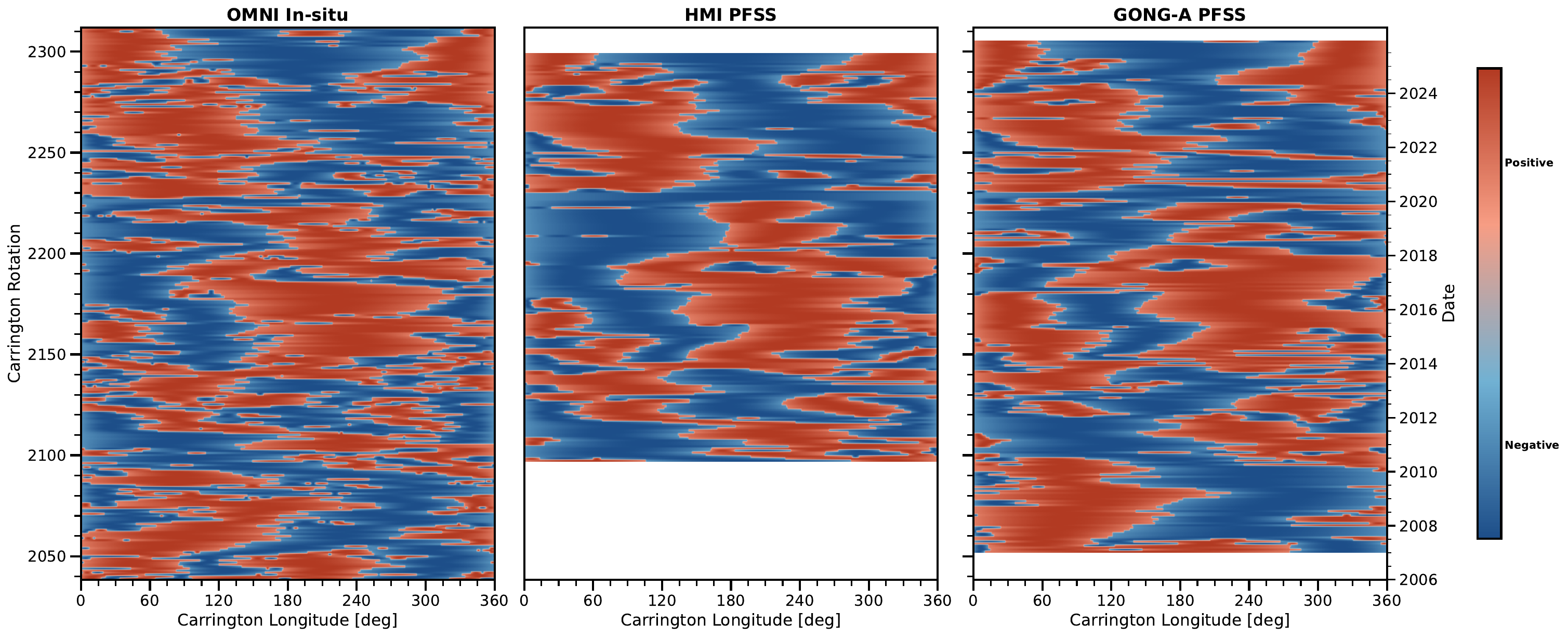}
   \caption{Stackplots of the magnetic field polarity as a function of Carrington longitude (x-axis) and time (y-axis) spanning 2007 through 2026. Panels show in-situ OMNI measurements (left), HMI PFSS model calculations at 1~au (middle), and GONG-A PFSS model calculations at 1~au (right). Color contours denote positive (red) and negative (blue) magnetic field polarities. For visual comparison, the PFSS data are shifted by $65^{\circ}$ to account for the nominal solar wind ($\approx 350\,\text{km\,s}^{-1}$) propagation delay to $1\,\text{au}$. Although the underlying polarity data are binary ($\pm 1$), continuous color shading is applied to enhance visual clarity.
   }
              \label{fig:imf_stack}%
\end{figure*} 

\begin{figure}[h]
   \centering
   \includegraphics[width=1\linewidth]{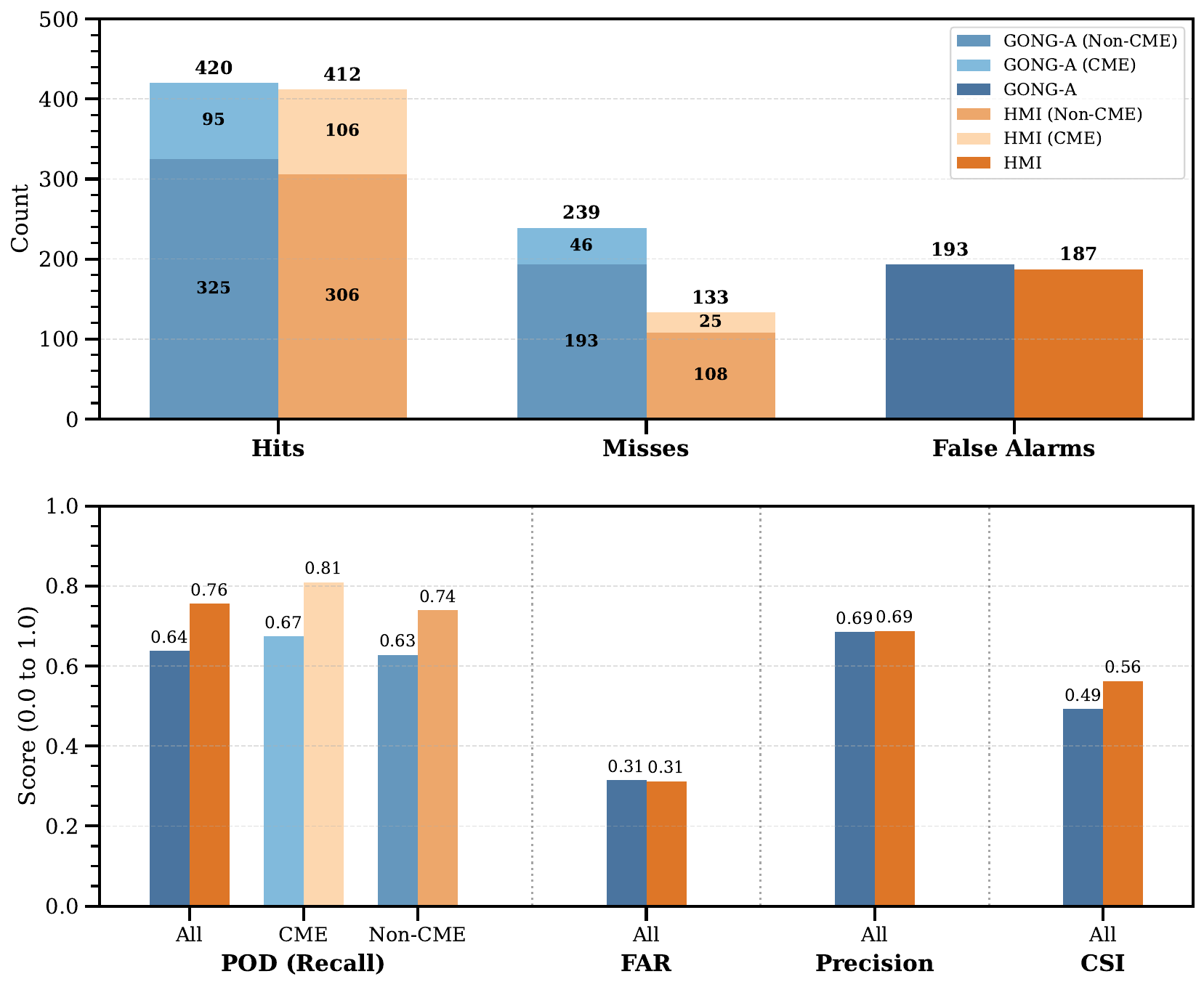}
   \caption{Hit-miss statistics and performance metrics for GONG-A and HMI datasets. The top panel shows absolute counts of Hits, Misses, and False Alarms categorized by CME association (Non-CME, CME, and All). The bottom panel displays the calculated skill scores: Probability of Detection (POD), False Alarm Ratio (FAR), Precision, and Critical Success Index (CSI) (see Section~\ref{subs:matrix}).}
              \label{fig:stats}%
\end{figure}

\begin{figure}[h]
   \centering
   \includegraphics[width=1\linewidth]{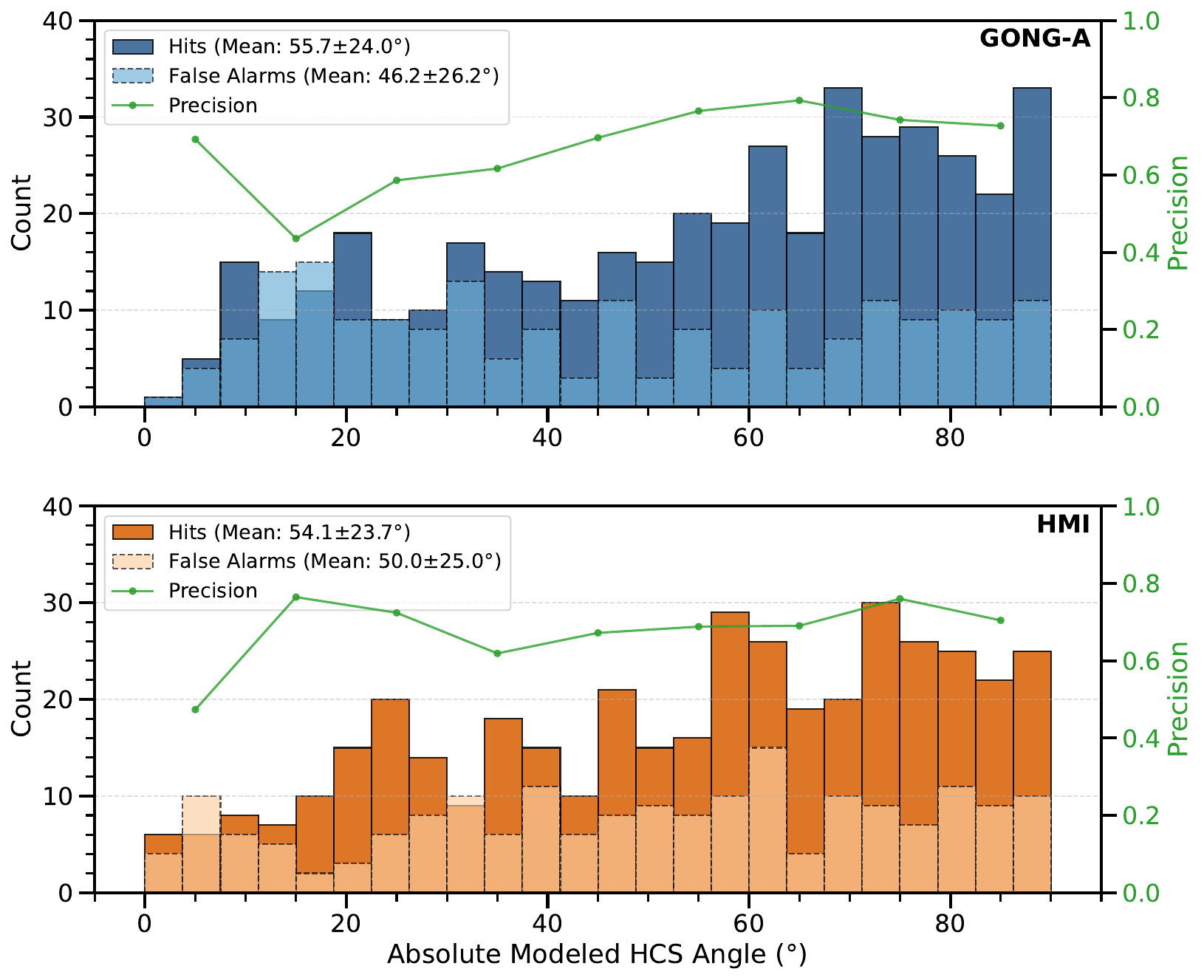}
    \caption{Distributions of hits (solid orange bars) and false alarms (dashed light-orange bars) alongside precision (green line) as a function of the absolute modeled HCS inclination angle for GONG-A (top) and HMI (bottom) data.}
              \label{fig:angle_dist}%
\end{figure} 

\begin{figure*}[h]
   \centering
   \includegraphics[width=1\linewidth]{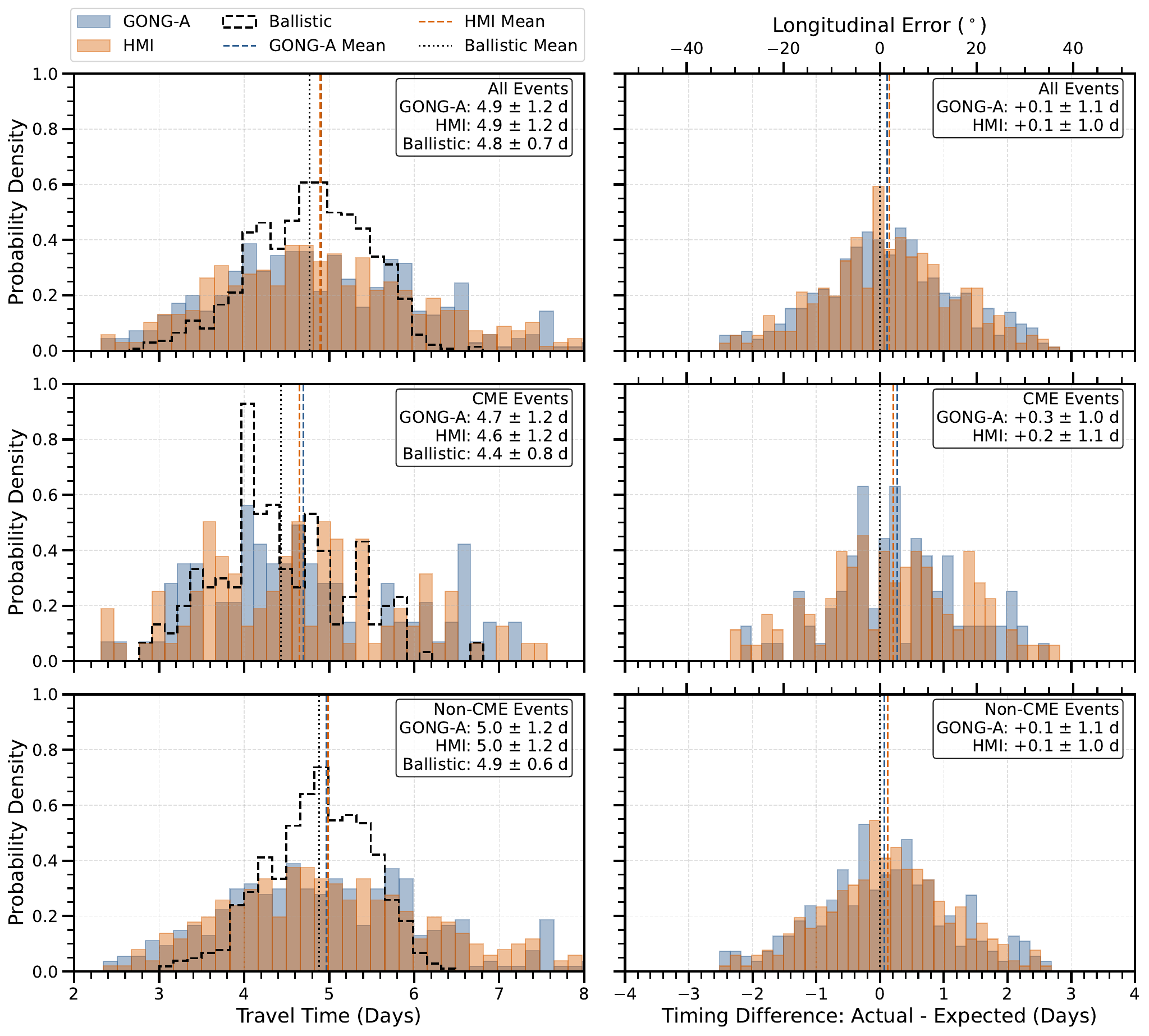}
   \caption{Travel time probability density distributions (left column) and timing difference distributions ($\tau_{\text{err}} = \tau_{\text{obs}} - \tau_{\text{exp}}$, right column) comparing GONG (blue), HMI (orange), and theoretical ballistic propagation (black dashed line). Rows categorize the dataset into all events (top), CME-associated events (middle), and non-CME-associated events (bottom). Mean values and standard deviations for each dataset are indicated by dashed vertical lines.}
              \label{fig:travel_time}%
\end{figure*}

\subsection{Data}\label{subs:data}
In this study, we employ two distinct magnetic map datasets to perform Potential Field Source Surface \citep[PFSS;][]{Altschuler1969} modeling of the solar corona, allowing us to determine the neutral line, and thus approximate the location of the HCS at a given source surface height. To model HCS crossings in the corona, we use two distinct full-disk magnetic field maps as lower boundary conditions. Using two independent datasets accounts for processing and observational uncertainties inherent to different data products. Furthermore, comparing daily updated maps with Carrington synoptic maps allows us to capture temporal uncertainties arising from differential rotation, far-side evolution, and short term flux emergence.\\ 

First, we utilize synoptic Carrington magnetic maps (\texttt{hmi.Synoptic\_Mr\_720s}) from the Helioseismic and Magnetic Imager \citep[HMI;][]{2012schou_HMI,2016couvidat_HMI} on board the Solar Dynamics Observatory \citep[SDO;][]{2012pesnell_SDO} provided by JSOC\footnote{Joint Science Operations Center; \url{http://jsoc.stanford.edu/}}. These maps are compiled by stacking full-disk observations from Earth's viewpoint across a complete solar rotation, yielding one map per Carrington rotation. The HMI dataset spans from 2010 to 2026. \\

Second, we incorporate flux transport magnetic maps from the Air Force Data Assimilation Photospheric Flux Transport model \citep[ADAPT;][]{2010Arge}, using Global Oscillations Network Group \citep[GONG;][]{Harvey1996_gong} magnetograms as input. The GONG-ADAPT maps (referred to as simply GONG-A in the further manuscript) are analyzed at a daily cadence and cover the timeframe from 2006 to 2026, using the first of the twelve ADAPT realizations and the map closest to 02:00 UTC (i.e., the first map of the day) each day to ensure a consistent and unbiased selection. \\

\subsection{Potential Field Extrapolation}

To reconstruct the coronal magnetic field, we utilize the PFSS code \texttt{CIDER}\footnote{\href{https://github.com/jpomoell/cider}{https://github.com/jpomoell/cider}} \citep[see, e.g.,][]{2024Heinemann_openflux}. The underlying potential solver discretizes Laplace's equation for the scalar potential on a staggered mesh via a finite-difference scheme. This discretization naturally preserves both $\nabla \cdot \mathbf{B} = 0$ and $\nabla \times \mathbf{B} = 0$ down to machine precision. Furthermore, exact preservation of the boundary conditions allows for rapid, high-resolution field extrapolation across the entire datasets.\\

The model is executed at a spatial resolution of $1^{\circ} \times 1^{\circ}$ in longitude and latitude, using 100 radial grid cells arranged in a stretched radial configuration. To match this resolution, input datasets are rebinned to $1^{\circ}$ using flux-conserving rebinning. This grid resolution accurately captures the large-scale coronal field geometry at the source surface height, which we place at $R_{\mathrm{ss}} = 2.5~\mathrm{R}_{\odot}$ in accordance with standard operational models \citep{McGregor2011}.\\

To identify modeled HCS crossings, we calculate the points of intersection between the polarity inversion line at $R_{\mathrm{ss}} = 2.5~\mathrm{R}_{\odot}$ and the projected orbital path of Earth. For each crossing event, we record the date, the polarity transition direction (outward-to-inward or vice versa), and the local angle at the intersection point of the modeled HCS and Earth's projected trajectory (see Appendix Fig.~\ref{fig:angle}). For the HMI dataset, all neutral line crossings are cataloged per Carrington rotation. For the GONG-A dataset, we consider only crossings falling within the central $14^{\circ}$ longitude band to prioritize the most newly updated magnetic field data at a 1 day cadence. Consecutive daily duplicate entries are filtered out. Figure~\ref{fig:pfss} illustrates representative cases for HMI (top) and GONG-A (bottom).

\subsection{Evaluation} \label{subs:eval}

To compare the modeled HCS crossings to in-situ measured ones, we use the publicly available catalog of sector boundary crossing SBCross\footnote{\href{https://sbcross.helsinki.fi}{https://sbcross.helsinki.fi}} \citep[][]{Heinemann2026_SBCross}. The maintained catalog includes sector boundary crossings since 1995 and their associated plasma and magnetic field parameters. Detection is based on an automated algorithm combining smoothed 255~eV suprathermal electron pitch angle distributions and the azimuthal angle of the interplanetary magnetic field to identify large-scale transitions. Building upon this database, our current study focuses specifically on these discrete sector boundary crossings rather than the extended sectors of continuous magnetic polarity in between. This choice is motivated by the fact that these transition regions contain dense, high-$\beta$ HCS and heliospheric plasma sheet structures that can strongly modulate and enhance the geoeffectiveness of solar transients \citep[e.g., ][]{Crooker2000,Temmer2025}.\\

For each in-situ event in the SBCross catalog starting from December 2006 (corresponding to the start of GONG-A data availability) observed at time $t_{\text{obs}}$ (which was chosen to be the center of the sector boundary crossing interval), we calculate the expected Sun-to-Earth transit time $\tau_{\text{exp}}$ assuming a constant-velocity ballistic propagation:
\begin{equation}
    \tau_{\text{exp}} = \frac{R_{\mathrm{1\,au}}}{v_{\text{mean}}},
\end{equation}
where $R_{\mathrm{1\,au}} = 1\,\text{au}$ and $v_{\text{mean}}$ is the observed mean plasma velocity of the sector boundary crossing. Although simple ballistic propagation ignores solar wind acceleration and non-radial flows (angular momentum) near the Sun, these two effects tend to oppose and cancel each other out, making the ballistic approximation perform well in practice \citep[][]{Macneil2022}.\\

To incorporate solar wind speed variability ($\pm v_{\text{std}}$) and non-ballistic acceleration effects, we establish lower and upper bounds on the transit time:
\begin{align}
    \tau_{\text{min}} &= \frac{R_{\mathrm{1\,au}}}{v_{\text{mean}} + v_{\text{std}}}, \\
    \tau_{\text{max}} &= \frac{R_{\mathrm{1\,au}}}{v_{\text{mean}} - v_{\text{std}}}.
\end{align}
The mean solar wind speed ($v_{\text{mean}}$) and its standard deviation ($v_{\text{std}}$) are taken directly from the SBCross catalog for the defined sector crossing. We further add a padding of $\Delta \tau = 2.5\,\text{days}$ to the maximum and minimum transit times ($\tau_{\text{min}}, \tau_{\text{max}}$) to cover the full spectrum of solar wind speeds and to account for model and propagation uncertainties such as stream interactions. The temporal boundaries for back-projecting to the solar source time are thus given by:
\begin{align}
    t_{\text{min}} &= t_{\text{obs}} - (\tau_{\text{max}} + \Delta \tau), \\
    t_{\text{max}} &= t_{\text{obs}} - \max\left(\tau_{\text{min}} - \Delta \tau,\, 2.31\,\text{days}\right),
\end{align}
where $2.31\,\text{days}$ is the absolute physical lower limit on travel time, corresponding to an upper solar wind propagation speed of $750\,\text{km\,s}^{-1}$ \citep[][]{Xu2015}. A PFSS-derived crossing is considered a successful match if it falls within $[t_{\text{min}}, t_{\text{max}}]$ and exhibits the same polarity reversal direction. When multiple candidate crossings exist within the window, the candidate closest to the expected crossing time ($t_{exp} = t_{obs} - \tau_{exp}$) is selected.\\

To investigate the influence of solar transients, we exploit the CME metadata tag included in the SBCross database, which marks sector boundary crossings occurring in close proximity to CMEs \citep[as given in the ongoing Richardson and Cane ICME catalog\footnote{\url{https://izw1.caltech.edu/ACE/ASC/DATA/level3/icmetable2.htm}};][]{2010richardson_RC-list}. Specifically, a crossing is flagged as CME-associated if its interval overlaps with an ICME passage or if the start/end times of an ICME fall within $\pm 12$ hours of the sector boundary crossing window. We can utilize this classification to assess whether interactions between CMEs and HCS structures systematically alter how well the model matches to the observations.

\section{Results}\label{s:res}

\subsection{Large Scale Structure} \label{subs:large-scale}
%To investigate the large-scale structure of the interplanetary magnetic field, we stacked time series of the magnetic polarity ($B_r$)  of each solar rotation as a function of time. 
%To investigate the large-scale structure of the interplanetary magnetic field, we stacked rotation-by-rotation time series of radial magnetic field polarity ($B_r$) extracted along Earth's trajectory as well as comparing in-situ OMNI measurements with PFSS model reconstructions."
%We did this for OMNI data at 1~au as well as PFSS data from HMI and GONG-A at $2.5\,R_{\odot}$. This is shown in Figure~\ref{fig:imf_stack}. We can see that the large-scale structure that was measured in-situ is well reproduced in the magnetic modeling results, although differences exist on smaller scales. Thus, we expect that PFSS can be used to model sector boundary crossings reasonably well.

To investigate the large-scale configuration of the interplanetary magnetic field, we stacked rotation-by-rotation time series of the radial magnetic field polarity ($\text{sign}(B_r)$) extracted along Earth's trajectory from both in-situ OMNI measurements at $1\,\text{au}$ and PFSS model reconstructions at $2.5\,R_{\odot}$ (Figure~\ref{fig:imf_stack}). A point-by-point comparison at $1^{\circ}$ resolution yields a baseline Jaccard index of $J \approx 0.33$, which peaks at $J \approx 0.50$ following a $65^{\circ}$ longitudinal shift. This offset corresponds to an average solar wind speed of approximately $350\,\text{km s}^{-1}$ and a 5-day travel time. Here, the Jaccard index is defined as
\begin{equation}
    J = \frac{\vert{}A \cap B\vert{}}{\vert{}A \cup B\vert{}}
\end{equation}
where $A$ represents the modeled polarity (at $r = 2.5\,R_{\odot}$; Fig.~\ref{fig:imf_stack}, middle or right panel) and $B$ represents the corresponding set of observed in-situ polarities at $1\,\text{au}$ (Fig.~\ref{fig:imf_stack}, left panel).

The moderate similarity score confirms that while minor small-scale discrepancies and temporal offsets exist, PFSS modeling accurately reproduces the overall sector topology and provides a reliable basis for identifying sector boundary crossings.

\subsection{Confusion Matrix} \label{subs:matrix}

Based on the evaluation methodology described in Section~\ref{subs:eval}, we performed a hit-miss statistical analysis on the detected HCS crossings, as illustrated in Figure~\ref{fig:stats} (results for solar minima and maxima can be found in the Appendix Fig.~\ref{fig:maxstats}, and \ref{fig:minstats}). We define the classification outcomes as follows:
\begin{itemize}
    \item {Hit ($TP$):} A modeled HCS crossing that successfully matches an observed in-situ event.
    \item {Miss ($FN$):} An observed in-situ HCS crossing without a corresponding modeled event.
    \item {False Alarm ($FP$):} A modeled HCS crossing with no corresponding in-situ event.
\end{itemize}

Hits and misses are further categorized into CME-associated and non-CME-associated events. Because CMEs are not included in the PFSS model but only derived from in-situ data, false alarms are evaluated solely against the total set of in-situ observations without CME categorization.

To quantify performance, we calculate the Probability of Detection ($\text{POD}$), False Alarm Ratio ($\text{FAR}$), Precision ($\text{PREC}$), and Critical Success Index ($\text{CSI}$) using the following standard formulations \citep[e.g., see][]{jolliffe2012forecast,wilks2019statistical}:
\begin{align}
    \text{POD} &= \frac{TP}{TP + FN} \label{eq:pod} \\[4pt]
    \text{FAR} &= \frac{FP}{TP + FP} \label{eq:far} \\[4pt]
    \text{PREC} &= \frac{TP}{TP + FP} = 1 - \text{FAR} \label{eq:prec} \\[4pt]
    \text{CSI} &= \frac{TP}{TP + FN + FP} \label{eq:csi}
\end{align}

Using GONG-A data between December 2006 and December 2025, we identified 613 modeled HCS crossings corresponding to 659 in-situ observed sector boundary crossing events. Of these, 420 were identified as hits, 239 as misses, and 193 as false alarms. Approximately $21\%$ of the in-situ events were associated with a CME; however, no statistically significant difference was observed between the CME and non-CME rates for hits and misses. This trend is reflected in the overall $\text{POD}$ of $0.64$, which shows a slightly higher value for CME-associated events ($0.67$) than for non-CME-associated events ($0.63$). The overall metrics for GONG-A data yield a $\text{FAR}$ of $0.31$, a $\text{PREC}$ of $0.69$, and a $\text{CSI}$ of $0.49$. \\

For the HMI dataset (evaluated from May 2010 to June 2025), 599 modeled crossings correspond to 545 in-situ events. This yields improved metrics across all parameters: a $\text{POD}$ of $0.76$, an $\text{FAR}$ of $0.31$, a $\text{PREC}$ of $0.69$, and a $\text{CSI}$ of $0.56$ (Figure~\ref{fig:stats}).\\

In the Appendix, Figures~\ref{fig:maxstats} and~\ref{fig:minstats} compare the hit-miss statistics during solar activity maxima (2010--2015, 2022--2026) and minima (2004--2009, 2016--2021). The $\text{POD}$ is notably higher during solar maxima ($\text{GONG-A} = 0.70$; $\text{HMI} = 0.81$) than during solar minima ($\text{GONG-A} = 0.56$; $\text{HMI} = 0.67$), again exhibiting slightly higher values for CME-associated events. $\text{CSI}$, $\text{FAR}$ and Precision show performance gains during solar maxima. We tested the evaluation using the in-situ measured start times of the sector boundary crossing intervals instead of the center points; this resulted in variations of 0.1-0.2 days in the timing difference and no significant change in the hit--miss results.\\

To evaluate algorithm quality and test for noise, data artifacts, or prediction bias, we established a randomized baseline of HCS crossings ($N = 550$, covering the same period as HMI). This yields a $\text{POD}$ of $\approx 0.2$, an $\text{FAR}$ of $\approx 0.8$, a $\text{PREC}$ of $\approx 0.2$, and a $\text{CSI}$ of $\approx 0.1$ (Appendix Fig.~\ref{fig:stats_rng}). All results are summarized in Table~\ref{tab:stats}.

\begin{table*}[htbp]
\centering
\begin{threeparttable}
\caption{Summary of the statistical results for GONG-A and HMI data including randomized baseline data (RNG). Data for the respective full time periods are shown in Figure~\ref{fig:stats}, and the randomized baseline data comparison is shown in Figure~\ref{fig:stats_rng} in the Appendix. Figures~\ref{fig:maxstats} and~\ref{fig:minstats} in the Appendix show the data during solar maxima and minima, respectively.}
\label{tab:stats}

\begin{tabular}{l|c|ccc|cc|cc}
\hline \hline
\textbf{Metric / Parameter} & \textbf{GONG} & \textbf{GONG} & \textbf{HMI} & \textbf{RNG} & \textbf{GONG} & \textbf{HMI} & \textbf{GONG} & \textbf{HMI} \\
\hline
\textbf{Time Periods} & 2006--2026 & \multicolumn{3}{c|}{2010--2026} & \multicolumn{2}{c|}{Maxima\textsuperscript{\textup{i} }} & \multicolumn{2}{c}{Minima\textsuperscript{\textup{ii} }} \\
\hline \hline
Hits ($TP$)        & 420 & 356 & 412 & 117 & 252 & 273 & 168 & 139 \\
\quad -- Non-CME  & 269 & 327 & 306 & 87  & 177 & 189 & 148 & 117 \\
\quad -- CME      & 95  & 87  & 106 & 30  & 75  & 84  & 20  & 22 \\
Misses ($FN$)      & 239 & 185 & 133 & 428 & 107 & 65  & 131 & 68 \\
\quad -- Non-CME  & 142 & 133 & 108 & 327 & 79  & 48  & 113 & 60 \\
\quad -- CME      & 46  & 43  & 25  & 101 & 28  & 17  & 18  & 8 \\
False Alarms ($FP$)& 193 & 158 & 187 & 433 & 105 & 121 & 88  & 66 \\
\hline
POD (Overall)     & 0.64 & 0.66 & 0.76 & 0.21 & 0.70 & 0.81 & 0.56 & 0.67 \\
POD (CME)         & 0.67 & 0.67 & 0.81 & 0.23 & 0.73 & 0.83 & 0.53 & 0.73 \\
POD (Non-CME)     & 0.65 & 0.65 & 0.74 & 0.21 & 0.69 & 0.80 & 0.57 & 0.66 \\
$\text{FAR}$      & 0.31 & 0.31 & 0.31 & 0.79 & 0.29 & 0.31 & 0.34 & 0.32 \\
Precision ($\text{PREC}$) & 0.69 & 0.69 & 0.69 & 0.21 & 0.71 & 0.69 & 0.66 & 0.68 \\
$\text{CSI}$      & 0.49 & 0.51 & 0.56 & 0.12 & 0.54 & 0.59 & 0.43 & 0.51 \\
\hline \hline
\end{tabular}
\begin{tablenotes}
\tiny
\item[\textup{i}] 2010--2015 and 2022--2026.
\item[\textup{ii}] 2004--2009 and 2016--2021.
\end{tablenotes}

\end{threeparttable}
\end{table*}

\subsection{Dependence on HCS Angle} \label{subs:matrix}

Using the PFSS model, we derived the local inclination angle of the HCS at its intersection with the projected trajectory of Earth (see Appendix Fig.~\ref{fig:angle}) as a proxy for its overall inclination and structural complexity. Figure~\ref{fig:angle_dist} presents the distributions of the absolute inclination angles for both hits and false alarms, along with precision as a function of angle for the HMI and GONG-A datasets. \\

The distribution of hits is skewed toward larger inclination angles, peaking at high inclinations ($\text{mean} \approx 56^\circ$ for GONG-A; $\text{mean} \approx 54^\circ$ for HMI). In contrast, false alarms exhibit a flatter distribution with a preference toward lower inclinations ($\text{mean} \approx 46^\circ$ for GONG-A; $\text{mean} \approx 50^\circ$ for HMI). This structural behavior is directly reflected in precision, which demonstrates a clear positive trend from low to high inclinations, a trend that is noticeably more pronounced in the GONG-A dataset than in the HMI dataset.

\subsection{Travel Time Distribution} \label{subs:travel_time}
We can investigate the propagation time of the HCS from the source surface to 1~au to identify possible modeling biases and deviations from simple ballistic propagation. Figure~\ref{fig:travel_time} shows the travel time distributions (first column) and  timing difference distribution(second column) for all correctly modeled events (hits), as well as separated into CME-associated and non-CME-associated events. \\

For the overall hit dataset, we find mean observed travel times $\overline{\tau_{\text{obs}}}$ of $4.9 \pm 1.2$~days for GONG-A and HMI. While the mean values closely match the expected ballistic travel time ($\overline{\tau_{\text{exp}}} = 4.8 \pm 0.7$~days), the observed distribution width is significantly broader than expected. The timing difference distribution, defined as $\tau_{\text{err}} = \tau_{\text{obs}} - \tau_{\text{exp}}$, is centered near zero ($\approx \pm 1.1$~days), with $90\%$ of the data falling within $[-39, +49]$~hours for GONG-A and $[-39, +44]$~hours for HMI. For larger timing differences ($>24$ hours) we find  a higher proportion of modeled low inclination angles (flat HCS) than for short timing differences (not shown). Further note that the timing difference can also be interpreted as longitudinal error in the PFSS model. 
For larger timing differences ($>24$\,hours), we observe a higher proportion of low modeled inclination angles (i.e., a flatter HCS) compared to events with smaller timing differences. Furthermore, these timing differences can be directly interpreted as longitudinal errors of the HCS crossing within the PFSS model results.\\

When differentiating between CME-associated and non-CME-associated events, CME-associated events exhibit a mean travel time that is $\approx 7$~hours shorter than their non-CME counterparts. The distribution widths remain similar ($\approx 1.2$~days). No significant differences are observed in the timing differences between these subsets; like the overall dataset, both display symmetric distributions with standard deviations of approximately $1$~day.\\

\section{Discussion}\label{s:disc}
Using PFSS modeling, we evaluated HCS crossing timings between 2006 and 2026 using GONG-A and HMI magnetic field data, comparing these outputs against in-situ sector boundary crossings from the SBCross catalog. Overall, the large-scale polarity structure is well reproduced (up to $J \approx 0.50$), with discrepancies primarily occurring on smaller temporal scales. This is consistent with previous studies: \cite{Wang2025} reported a correlation of 0.55--0.60 between PFSS-derived magnetic sectors and in-situ observations; \cite{Hoeksema1982,1983Hoeksema} demonstrated a $80\%$ daily polarity match when employing $R_{\text{ss}} = 2.5\,R_{\odot}$ with $10\,\text{G}$ sharply defined poles; and \cite{Hoeksema1995} found a correlation of 0.59.\\

In our analysis, the model achieved overall hit rates of 0.64 and 0.76 for GONG-A and HMI data, respectively. We note that standard Carrington synoptic maps (HMI) performed slightly ($10$--$20\%$) better at modeling HCS crossings than daily updated flux-transport maps (GONG-A). While newly assimilated data minimizes local aging, flux transport alters the broader surface field distribution, which can shift the global polarity inversion line. We hypothesize that daily assimilated magnetograms introduce short-term spatial fluctuations into the modeled HCS that degrade matching between model and observation more than the temporal aging inherent to classic Carrington maps \citep[][]{Heinemann2021_farside,Posner2021}. We observed a hit-rate variation of approximately 0.14 between solar maximum and solar minimum, with the model performing significantly better during phases of high solar activity. This aligns with \cite{Wang2025}, who also identified improved polarity matching during higher activity phases, which is contrary to earlier studies that reported better performance during quiet solar conditions \citep[e.g.,][]{Jian2015_CCMC}. We attribute these results to the evolving geometric tilt and configuration of the HCS over the solar cycle \citep[][]{Suess1993}. Specifically, we find that higher HCS inclination angles correspond to a significantly improved hit rate, as the wavy structure characteristic of solar maximum increases the likelihood of Earth intersecting sector boundaries. Conversely, at larger timing differences ($>24$\,h), we observe a higher proportion of lower inclination angles. During solar minimum, when the HCS adopts a flatter, low-inclination profile, even minor uncertainties in modeled latitude can determine whether a given sector boundary crossing is successfully detected at 1\,au.\\

Interestingly, interactions between CMEs and HCS structures did not degrade the model's hit rate. Instead, we observed a minor (likely non-significant) increase in performance. This contrasts with expected behavior, as HCS interactions are known to introduce substantial travel-time uncertainties into CME propagation forecasts \citep[e.g.,][]{Kay2015a,Winslowetal2016}. Although CME-associated HCS events exhibited travel times up to 7~hours shorter due to elevated average solar wind speeds, these typically steep HCS crossings were still reliably detected in-situ. CMEs may fail to degrade the overall hit rate because the global HCS rapidly reforms post-interaction \citep[][]{Zhao1996,Blanco2011}, or because CMEs are channeled along the HCS structure itself \citep[e.g.,][]{Crooker1996, Temmer2026}. While such interactions can alter the local in-situ inclination of the HCS \citep[e.g.,][]{Pizzo1991}, detailed local orientation changes were beyond the scope of this study.\\

Our analysis demonstrates a travel-time spread of up to $\approx \pm 2$~days, which is comparable to typical arrival-time errors in operational space weather forecasting for solar transients \citep[][]{Temmer2021,Temmer2023} and high-speed solar wind streams \citep[e.g, see][]{gressl14,Reiss2016,Hinterreiter2019}\\

\section{Summary and Conclusions}\label{s:sum}

In this study, we evaluated the capability of static PFSS extrapolations ($R_{\mathrm{ss}} = 2.5\,R_{\odot}$) to reproduce $1\,\mathrm{au}$ sector boundary crossings identified in the SBCross catalog between 2006 and 2026. Rather than employing complex solar wind propagation models, we analyze the static extrapolations directly to assess how accurately they capture large-scale heliospheric magnetic structure and track its evolution across the solar cycle. The main conclusions of this work are summarized as follows:

\begin{enumerate}
    \item Direct ballistic mapping of PFSS modeled HCS crossings to $1\,\mathrm{au}$ yields high detection rates, achieving POD scores of $0.64$ for GONG-ADAPT and $0.76$ for HMI synoptic data.  The standard Carrington synoptic maps perform about $10\text{--}20\%$ better than daily updated flux-transport maps. This indicates that short-term assimilation noise and missing far-side flux balance degrade topological matching more than the temporal aging of Carrington maps.

    \item  Model performance exhibits a dependence on solar activity, performing significantly better during solar maxima ($\mathrm{POD} = 0.70\text{--}0.81$, $\mathrm{CSI} = 0.54\text{--}0.59$) than during solar minima ($\mathrm{POD} = 0.56\text{--}0.67$, $\mathrm{CSI} = 0.43\text{--}0.51$). This variation is attributed to solar cycle variation in the tilt of the HCS: highly inclined current sheets during solar maximum are robustly intersected by Earth, whereas flat configurations near solar minimum are sensitive to uncertainties in latitude and solar wind propagation effects. This is consistent with the fact that the hit distributions are strongly skewed toward larger inclination angles ($\mathrm{mean} \approx 54^{\circ}\text{--}56^{\circ}$).
    
    \item CMEs do not degrade the model's ability to match sector boundary crossings ($\mathrm{POD} = 0.67\text{--}0.83$ for CME-associated SBCs). Although CME-associated events exhibit travel times approximately $7\,\mathrm{hours}$ shorter due to higher solar wind speeds, the global topological structure of the HCS remains intact or rapidly relaxes post-passage, allowing reliable identification.
    
    \item Observed travel times closely resemble ballistic propagation estimates ($\tau_{\mathrm{obs}} \approx 4.9 \pm 1.2\,\mathrm{days}$ vs. $\tau_{\mathrm{exp}} \approx 4.8 \pm 0.7\,\mathrm{days}$). The timing difference distribution ($\tau_{\mathrm{err}} = \tau_{\mathrm{obs}} - \tau_{\mathrm{exp}}$) is centered near zero ($ \pm 1.1\,\mathrm{days}$), with $90\%$ of matched events falling within a temporal uncertainty window of approximately $\pm 1.8\,\mathrm{days}$ ($\approx \pm 44\text{--}49\,\mathrm{hours}$).
\end{enumerate}

Overall, PFSS extrapolations provide an accurate, computationally inexpensive method for approximating the large-scale sector boundary crossings and interplanetary magnetic field polarities. However, capturing localized current sheet deformations, non-radial deflections, and the complex stream interactions driven by solar transients still requires time-dependent modeling \citep[e.g.,][]{Samara2025}.

%% Please use the acknowledgment and contribution environments. This will 
%% be anonomyized when the "anonymous" style option is used. 
\begin{acknowledgments}
 We acknowledge use of NASA/GSFC's Space Physics Data Facility's OMNIWeb/CDAWeb service, and OMNI data. SDO data are courtesy of NASA and the respective science teams. GONG data are provided by the National Solar Observatory, operated by the Association of Universities for Research in Astronomy (AURA) under a cooperative agreement with the National Science Foundation. SGH acknowledges funding from the Research Council of Finland (Academy Fellowship) [370747; RIB-Wind]. SGH thanks his chair for always having his back. This research was funded in whole or in part by the Austrian Science Fund (FWF) [10.55776/PAT2280725]. MO is  part-funded by Science and Technology Facilities Council (STFC) grant number UKRI1207 and Natural Environment Research Council (NERC) grant number NE/Y001052/1. MA-L acknowledges funding from the Research Council of Finland (Academy Fellowship) [376431; SICS] as well as the NSF Grant 2420675. AW and EK acknowledge the SolMAG project (ERC-COG 724391) funded by the European Research Council (ERC) in the framework of the Horizon 2020 Research and Innovation Programme. JP acknowledges the Research Council of Finland projects [364852, 370793; ENERGISE and 375401; MesoCor]. CIDER model development has received funding from the European Union's Horizon Europe research and innovation programme under grant agreement No 101134999 (SOLER). EK acknowledges ERC Synergy project GERACLE (4100617), funded by the EU under grant agreement 101166910. SGH, JP, EK MA-L and AW acknowledge the Research Council of Finland Centre of Excellence project SpaceResilience [1374096]. Open access funded by Helsinki University Library.
\end{acknowledgments}

\begin{contribution}
%%This section gives authors the space to recognize author contributions. The text inside this environment is NOT counted towards the total word quanta. At a minimum, manuscripts are expected to include this text:

SGH conducted the analysis and led the manuscript preparation. MT contributed the core concept for this study. JP provided the PFSS model. All authors participated in revising and editing the manuscript. 
%All authors contributed equally to the Terra Mater collaboration.

%% But authors are expected to provide more specific details, e.g. 
%%
%%SC was responsible for writing and submitting the manuscript.
%%WWM came up with the initial research concept and edited the manuscript.
%%OTS obtained the funding and edited the manuscript.
%%EBF provided the formal analysis and validation. He also edited the manuscript.
%%GEH Supervised the undergraduates, wrote the software and administers the project github and Zenodo repositories.
%%
%% Authors can use the Contributor Role Taxonomy (CRediT) at
%% for ideas on how write a good statement tailored to their needs.

\end{contribution}

%\facilities{HST(STIS), Swift(XRT and UVOT), AAVSO, CTIO:1.3m, CTIO:1.5m, CXO}

%\software{astropy %\citep{2013A&A...558A..33A,2018AJ....156..123A,2022ApJ...935..167A},  
   %       Cloudy \citep{2013RMxAA..49..137F}, 
   %       Source Extractor \citep{1996A&AS..117..393B}
   %       }


\appendix
\section{Additional Figures}
\begin{figure}[h]
   \centering
   \includegraphics[width=0.5\linewidth]{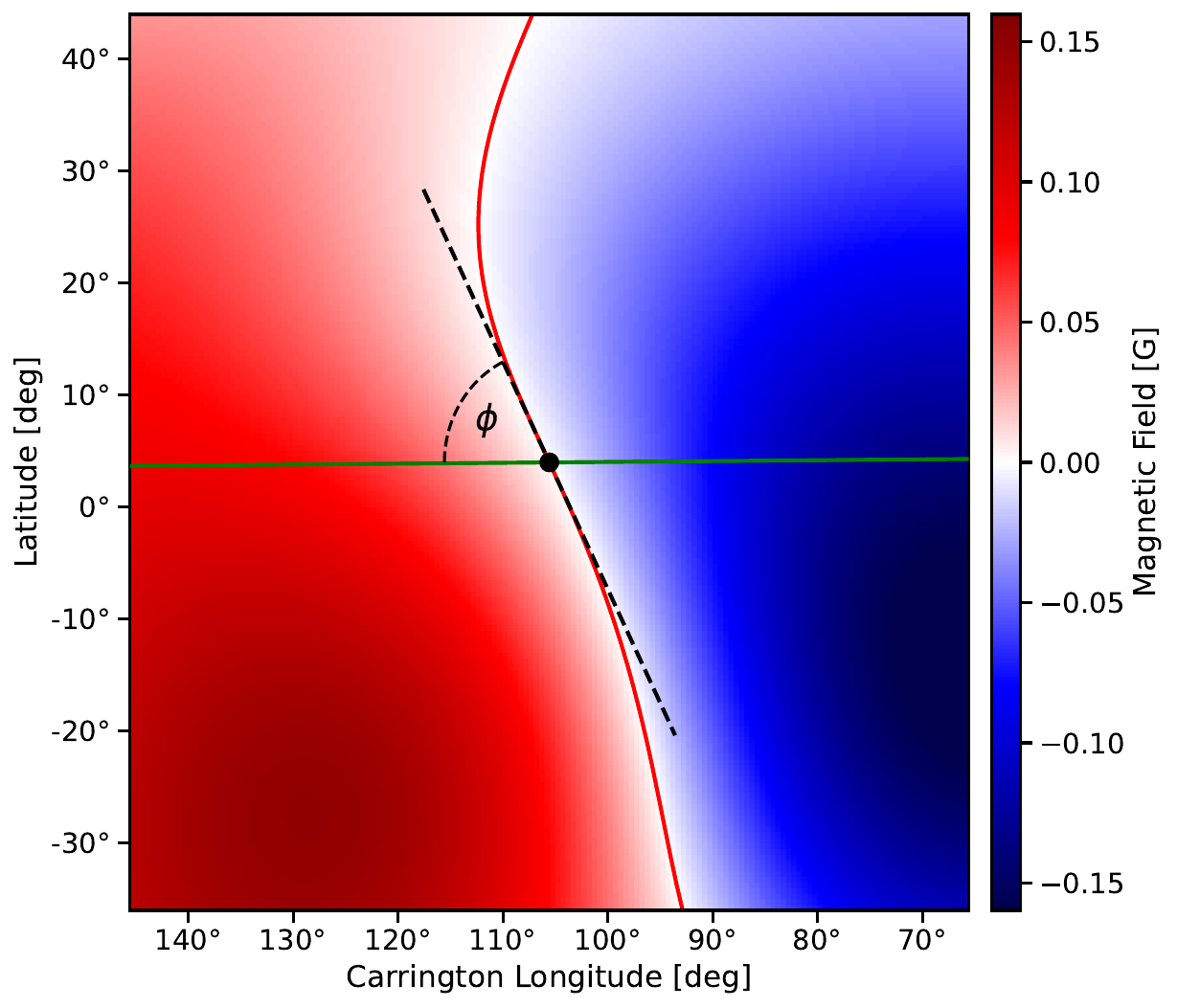}
   \caption{Schematic diagram illustrating the definition of the HCS inclination angle $\phi$ at the Earth's path intersection. The plot shows a region of the source surface map with the radial magnetic field strength, the HCS (red line), the projected trajectory of Earth (green line), and the modeled HCS crossing (black dot). The angle $\phi$ is measured between the HCS (dashed black line) and the projected Earth's trajectory.}
              \label{fig:angle}%
\end{figure}

\begin{figure}[h]
   \centering
   \includegraphics[width=1\linewidth]{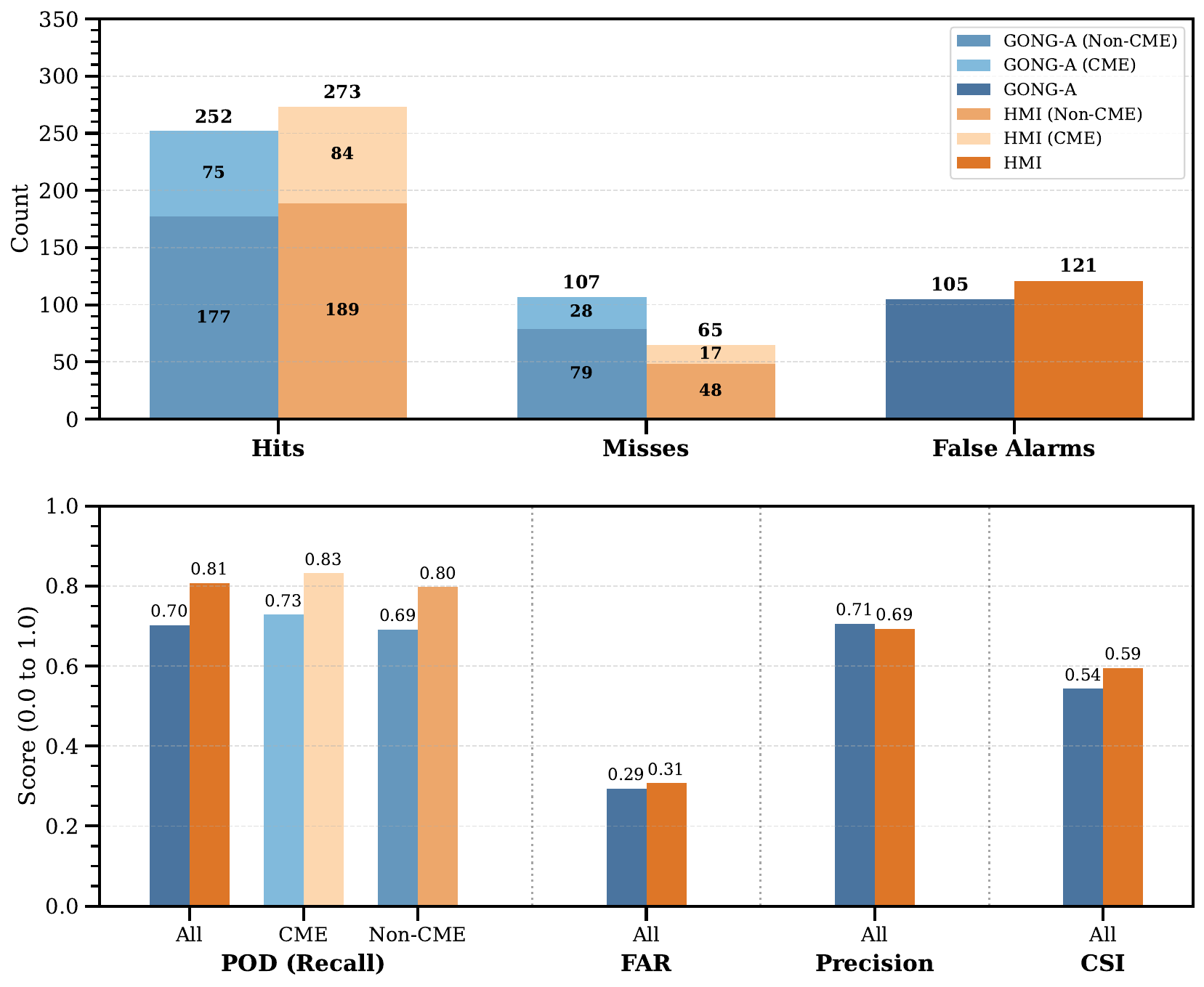}
   \caption{Same as Figure~\ref{fig:stats} but only for data in solar maxima (2010--2015, 2022--2026), highlighting model skill scores and contingency metrics during periods of increased magnetic complexity and frequent transients.}
              \label{fig:maxstats}%
\end{figure} 

\begin{figure}[h]
   \centering
   \includegraphics[width=1\linewidth]{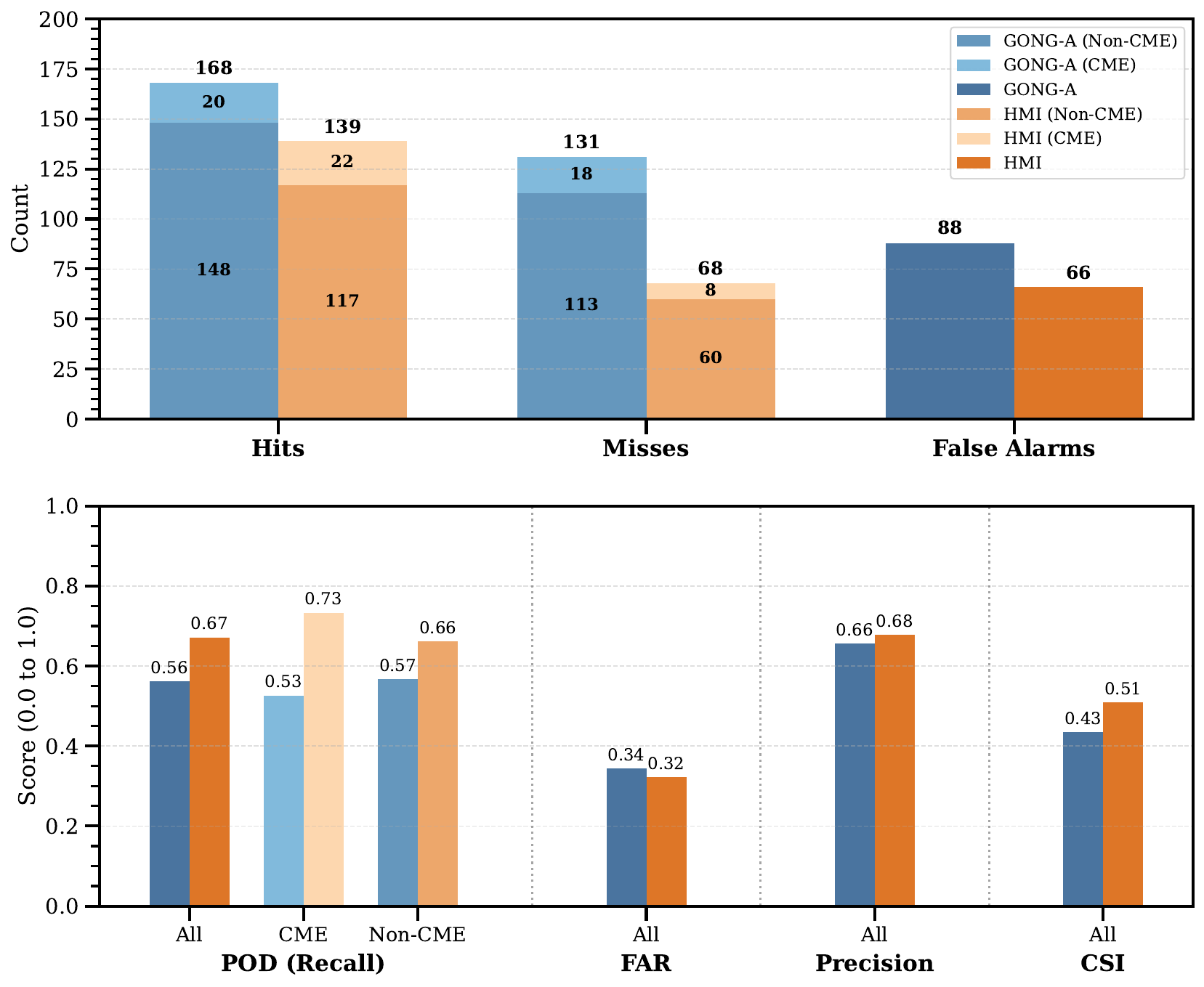}
   \caption{Same as Figure~\ref{fig:stats} but only for data in solar minima (2004--2009, 2016--2021), illustrating model performance and hit-miss statistics during periods of lower solar activity and a more stable magnetic dipole configuration.}
              \label{fig:minstats}%
\end{figure} 

\begin{figure}[h]
   \centering
   \includegraphics[width=1\linewidth]{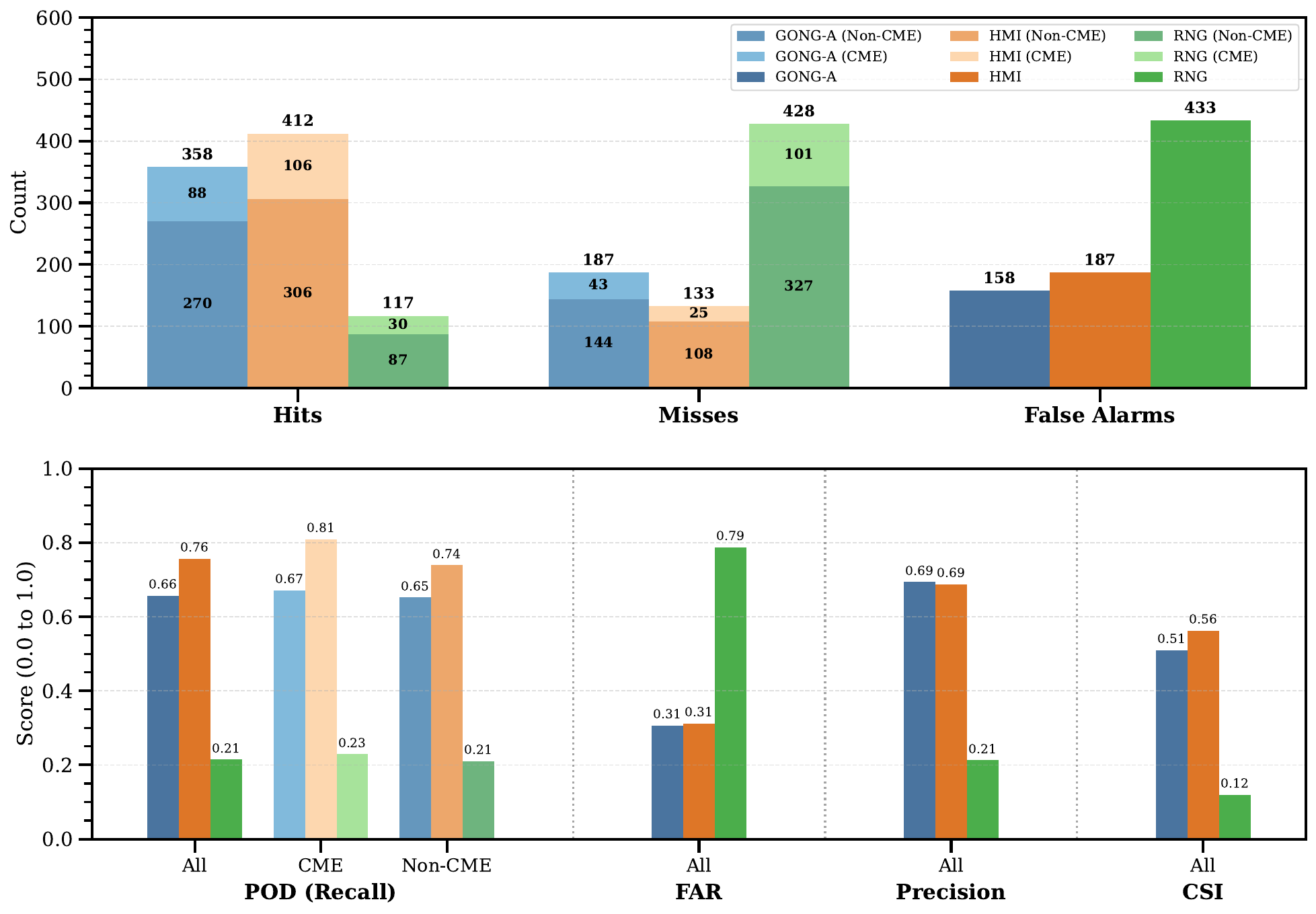}
   \caption{Same as Figure~\ref{fig:stats}, but restricted to the 2010--2026 period and including random predictions for HCS crossings (green), serving as a benchmark baseline to evaluate whether the model predictions perform significantly better than random chance.}
              \label{fig:stats_rng}%
\end{figure} 

%\begin{figure}[h]
 %  \centering
 %  \includegraphics[width=1\linewidth]{angles_timingdiff.pdf}
 %  \caption{angle and speed distribution for various timing difference %bins. \textcolor{red}{THIS FIGURE DOES NOT GO IN THE PAPER. ITs for %your convenience to look at.}}
%              \label{fig:no-in-paper}%
%\end{figure} 
%\include{table1}

%% For this sample we use BibTeX plus aasjournalv7.bst to generate the

%\bibliography{bib}{}
\bibliographystyle{aasjournalv7}



\end{document}